\documentclass [preprint]{cjaa}          
\usepackage{amsmath}
\usepackage{graphicx}
\begin{document}

   \title{Note on Redshift Distortion in Fourier Space}
   \volnopage{Vol.0 (200x) No.0, 000--000}   
   \setcounter{page}{1}         

   \author{Yan-Chuan Cai 
      \inst{1,2}\mailto{}
      \and Jun Pan
      \inst{1}
        }
   \offprints{Yan-chuan Cai}                 
   \institute{Purple Mountain Observatory, Chinese Academy of Sciences, 
              Nanjing 210008, China\\
             \email{cyc@bao.ac.cn}
             \and
             National Astronomical Observatories, Chinese Academy of Sciences,
             Beijing 100012, China\\
             }
   \date{Received~~2006 month day; accepted~~2006~~month day}

   \abstract{We explore features of redshift distortion in Fourier analysis of 
N-body simulations. The phases of the Fourier modes of the dark matter density
fluctuation are generally shifted by the peculiar motion along the line of sight,
the induced phase shift is stochastic and has probability distribution function (PDF)
symmetric to the peak at zero shift while the exact shape depends on the wave vector, 
except on very large scales where phases are invariant by linear perturbation theory.
Analysis of the phase shifts motivates our phenomenological models for the bispectrum 
in redshift space. Comparison with simulations shows that our toy models are
very successful in modeling bispectrum of equilateral and isosceles triangles at large 
scales. In the second part we compare the monopole of the power spectrum and bispectrum 
in the radial and plane-parallel distortion to test the plane-parallel approximation. 
We confirm the results of Scoccimarro (2000) that difference of power spectrum is at the 
level of $10\%$, in the reduced bispectrum such difference is as small as a few percents. 
However, on the plane perpendicular to the line of sight of $k_z=0$, the 
difference in power spectrum between the radial and plane-parallel approximation can be more 
than $\sim 10\%$, and even worse on very small scales. Such difference is prominent for 
bispectrum, especially for those configurations of tilted triangles. The non-Gaussian 
signals under radial distortion on small scales are systematically biased downside than 
that in plane-parallel approximation, while amplitudes of differences depend on the opening 
angle of the sample to the observer. The observation gives warning to the practice of using 
the power spectrum and bispectrum measured on the $k_z=0$ plane as estimation of the real space
statistics.
   \keywords{cosmology: theory --- large-scale structure of universe --- methods: statistical}
   }
   \authorrunning{Yan-chuan Cai \& Jun Pan}
   \titlerunning{Note on Redshift Distortion in Fourier Space} 
   \maketitle

\section{Introduction}           

Large redshift surveys of galaxies provide us with distances estimated from measured 
redshifts to complete our three-dimensional topography of the visible universe. 
However, due to peculiar velocities of galaxies, the inferred redshift from observation 
is not an exact indication of the true distance, therefore our estimation of statistics 
in redshift space suffers from redshift distortion. On large scales 
peculiar velocities are dominated by bulk inflow, the redshift distortion 
boosts the clustering strength of galaxies, on small scales the virialised peculiar 
velocities suppress the clustering power and account for the 
phenomenon of ``Finger of God'' (e.g. Kaiser 1987; Hamilton 1998).

Redshift distortion is entangled with bias and the gravitational nonlinearity. 
Nevertheless, fruitful results of redshift distortion have been achieved 
and applied in statistical analysis of LSS, particularly the 
power spectrum $P(k)$ and the two point correlation function $\xi$. On large scales 
where linear theory applies, the power spectrum in redshift space is
\begin{equation}
P_s({\bf k})=(1+\beta \mu^2)^2 P(k)\ ,
\end{equation}
where $\beta=\Omega_m^{0.6}/b$ with $b$ the linear bias parameter, and 
$\mu$ is the cosine of the angle between ${\bf k}$ and the unit vector $\hat{z}$ of 
line-of-sight (LOS) (Kaiser 1987). 
Further observation and analysis of N-body simulation denotes that over wide scales 
including strong nonlinear regime there is an accurate empirical formula to model the 
effect of redshift distortion on power spectrum with introduction of a pairwise velocity 
dispersion parameter $\sigma_v$ 
(e.g. Park et al. 1994; Peacock \& Dodds 1994; Cole, Fisher \& Weinberg 1995; 
Hatton \& Cole 1998; White 2001),
\begin{equation}
P_s({\bf k})=P(k)\frac{(1+\beta\mu^2)^2}{[1+(k\mu\sigma_v)^2/2]^2}\ .
\end{equation}

In comparison, understanding of redshift distortions for bispectrum or the three point 
correlation function is not satisfactory (Hivon et al. 1995; Verde et al. 1998; 
Scoccimarro et al. 1999), which, in part, is attributed to the non-perturbative nature 
of redshift distortion effect (Scoccimarro et al. 1999). 
The intrinsic difference between power spectrum and bispectrum is that 
power spectrum is a phase invariant quantity which contains only the information of the 
amplitude of the Fourier transform of the density contrast, bispectrum consists of 
contributions from both amplitudes and phases (Matsubara 2003). Research based on phases 
has revealed that phase information is very crucial in LSS distribution pattern 
recognition (Chiang \& Coles 2000; Chiang 2001; Chiang, Coles \& Naselsky 2002), 
especially for the bispectrum and the three point correlation function (Watts \& Coles 2003; 
Watts, Coles \& Melott 2003; Matsubara 2003; Hikage, Matsubara \& Suto 2004; Chiang 2004; 
Hikage et al. 2005). Therefore there is strong reason to believe that the behavior of phases
under redshift distortion will provide us with precious information, which is 
the starting point of this paper.

An alternative route to tackle redshift distortion is to recover real space quantities from
measurements in redshift space. The anisotropic two point correlation function in redshift
space $\xi({\bf s})$ is measured along two separations as $\xi(\sigma, \pi)$, where 
$\pi={\bf s}\cdot\hat{z}/s$ and $\sigma=|{\bf s}\times \hat{z}|/s$. The projected 
two point correlation function $\Xi(\sigma)$ is free of redshift distortion and simply 
related to the real space $\xi(r)$ by (Davis \& Peebles 1983; Fisher et al. 1994)
\begin{equation}
\Xi(\sigma)\equiv \int_{-\infty}^{+\infty} \xi(\sigma, \pi)d\pi=
2 \int_{\sigma}^{+\infty} \frac{r \xi(r) dr }{\left( r^2 - \sigma^2 \right)^{\frac{1}{2}}} \ ,
\end{equation}
or, we can deproject $\Xi$ to have the real space two point correlation function 
(e.g. Saunders, Rowan-Robinson \& Lawrence 1992; Hawkins et al. 2003)
\begin{equation}
\xi(r)=-\frac{1}{\pi}\int_r^{+\infty}\frac{d\Xi(\sigma)/d\sigma}{\left( \sigma^2-r^2 \right)^{\frac{1}{2}}} 
d\sigma \ .
\end{equation}
In Fourier space it is much simpler, as we can see from Eq.~1 and Eq.~2 that if we restrict 
our measurement on the plane of $k_z=0$ only, what we obtain is
the real space power spectrum (e.g. Park et al. 1994; Hamilton \& Tegmark 2002). 

However, the approach relies on the assumption of the parallel approximation, i.e. the 
observer is at infinite distance so the unit vector $\hat{z}$ of the LOS is pointing to a 
fixed direction. In reality there is no such fixed direction of LOS, systematic biases are 
inevitably introduced in the treatment mentioned above. Particularly since Fourier transform 
mixes all scales together, even at high $k$ there is some contamination from scales at wide 
angles. Consequently not only are polyspectra measured on 
$k_z=0$ plane not equivalent to those in real space, but also the monopoles of polyspectra 
with respect to LOS are not strictly following what theories of redshift distortion under 
plane approximation predict. Further complication also arises that Fourier modes are no 
longer independent and the covariance matrix is not diagonal (Zaroubi \& Hoffman 1996, 
Scoccimarro 2000).

Attention has been called in works analyzing galaxy redshift surveys which always subtend 
wide angles over sky and are at low redshift, to reduce the systematics in the adoption of 
parallel approximation. Numerical studies have shown that if one confines work with small 
angle, differences brought forward by parallel approximation in monopoles are negligible, 
in quadruples of power spectrum are usually under $\sim 5\%$ if the opening angle of the 
window in which Fourier transform is performed is less than $50^\circ$ 
(e.g. Park et al. 1994; Cole, Fisher \& Weinberg 1994). Scoccimarro (2000) also reported 
that such approximation was good enough for measuring the monopoles of the
power spectrum and the bispectrum. Even for all sky surveys the difference is as small 
as $10\%$. Whatsoever, as the parallel approximation plays a fundamental role in Fourier 
analysis of galaxies' spatial distribution,  it definitely deserves exclusive studies, 
which is very crucial in keeping confidence on our interpretation of 
what we estimated in redshift space. We will show in this paper that the real life is 
never simple.

Our paper is organized as follows. In Section 2, we give general description of mapping 
from real space to redshift space, demonstrate the shift of phases of Fourier modes 
by redshift distortion and discuss possible construction of empirical model for bispectrum 
in redshift space. Section 3 focuses on the difference between the plane-parallel approximation
and the radial distortion. Conclusion and discussion is in section 4. Throughout the paper 
we are using two sets of $\Lambda$CDM simulations by the Virgo consortium, the Virgo 
simulation in box of size $239.5h^{-1}$Mpc with $256^3$ particles and the VLS simulation 
in box of size $479h^{-1}$Mpc with $512^3$ particles, cosmological parameters
of the two simulations are $\Omega_m=0.3$, $\Omega_\Lambda=0.7$, $\sigma_8=0.9$ and 
$\Gamma=0.21$ (Jenkins et al. 1998).

\section{Phase information and bispectrum in redshift space}
\subsection{Phase shift is a nonlinear effect}
For simplicity in this section we adopt the plane approximation so the unit vector of 
LOS $\hat{z}$ has fixed direction. The mapping from real space coordinates ${\bf{r}}$ 
to redshift space $\bf{s}$ is
\begin{equation} 
\bf{s}=\bf{r} - \beta \bf{u}_z(\bf{r})\hat{z}\ , 
\end{equation}
where $\bf{u}(\bf{r}) \equiv - \bf{v}(\bf{r})/({\cal H} \beta)$, $\bf{v}(\bf{r})$ is the 
peculiar velocity, and ${\cal H}(\tau) \equiv (1/a)(da/d\tau)= Ha$ is the conformal Hubble 
parameter with FRW scale factor $a$ and conformal time $\tau$ 
(Scoccimarro et al. 1999; Scoccimarro 2004).

The density contrast in redshift space, $\delta_s(s)$, is obtained from the
real space density fluctuation $\delta(r)$ by requiring conservation of the number of 
galaxies,
\begin{equation} 
(1+\delta_s)d^3s=(1+\delta)d^3r
\end{equation}
Fourier transform of the number density contrast in redshift space reads
\begin{equation}
\delta_s(\bf{k}) \equiv \int \frac{d^3s}{(2\pi)^3} {\rm e}^{-{\it i} \bf{k}\cdot \bf{s}} \delta_s(\bf{s})
= \int  \frac{d^3x}{(2\pi)^3} {\rm e}^{-{\it i} \bf{k}\cdot \bf{x}}
{\rm e}^{{\it i} \beta k_z u(x)} \left[ \delta(\bf{x}) + \beta \nabla_z u_z(x) \right]\ ,
\end{equation}
in which and hereafter the subscript ``s'' refers to quantities in redshift space.
The term in square brackets describes
the squashing effect, {\it i.e.}, the boost to the clustering
amplitude, whereas the exponential factor encodes the
``Finger of God'' effect, which erases power due to velocity dispersion along the
LOS. In plane-parallel approximation, the distortion comes from 
the peculiar velocity along the LOS $v_z$. Thus, to generate a redshift space sample, we only 
need $s_z=r_z-\beta u_z$.

It is clear from Eq.~7 we see that in the redshift space, not only the amplitude 
$|\delta(k)|$, but also the phase angle $\theta(k)$ is distorted which is casually ignored. 
The expression can be expanded perturbatively as demonstrated by Soccimarro et al. (1999) 
who calculated the expansion to second order to model the bispectrum in redshift space on 
large scales. We reproduce here the expansion
\begin{equation}
\delta_s({\bf k})=\sum_{n=1}^\infty \int d^3 k_1 ... d^3 k_n [\delta_D]_n 
\left[ \delta({\bf k}_1)+\beta\mu_1^2\phi({\bf k}_1)\right]
\frac{(\beta\mu k)^{n-1}}{(n-1)!}\frac{\mu_2}{k_2}\phi({\bf k}_2)...\frac{\mu_n}{k_n}\phi({\bf k}_n)\ ,
\end{equation}
where $[\delta_D]_n=\delta_D(\bf{k}-\bf{k}_1-...-{\bf k}_n)$ is the Dirac function, 
velocity divergence is $\phi({\bf r})\equiv \nabla \cdot {\bf u}$ and $\mu_n={\bf k}_n\cdot \hat{z}/k_n$. 
In the linear theory only the first order $n=1$ counts, we recover the well-known Kaiser 
formula (Kaiser 1987)
\begin{equation}
\delta_s({\bf k})=(1+\beta \mu^2) \delta({\bf k})\ ,
\end{equation}
which tells that at the first order phases are not changed by redshift distortion, phase shift
is a phenomenon at high orders $n \ge 2$, i.e. nonlinear. We are not going to calculate 
quantitatively the phase shift in perturbation theory, rather, as a starting point in this 
direction, we aim at presenting the existence of the phase shift in numerical simulations and 
henceforth implication drawn from the information to understand redshift distortion on 
phase-related statistics, the bispectrum.

\subsection{Bispectrum and phase information}
By definition, the bispectrum is the ensemble average of the product of three Fourier modes 
with their wave vectors forming a triangle,
\begin{equation}
B({\bf k}_1, {\bf k}_2, {\bf k}_3=-{\bf k}_1-{\bf k}_2)=\langle \delta({\bf k}_1) \delta({\bf k}_2) 
\delta^*({\bf k}_3)\rangle\ = \langle |\delta_1| |\delta_2| |\delta_3| 
e^{{\it i} (\theta_1+\theta_2-\theta_3)} \rangle\ ,
\end{equation}
where we wrote the Fourier mode $\delta({\bf k})=|\delta|e^{{\it i}\theta}$. It is well 
known that the nonlinearity of gravitational evolution generates correlation between 
the phase $\theta$ and the amplitude $|\delta|$ as well as the correlations between different modes, 
especially on small scales although in the initial fluctuation there are no such correlations. 
Whilst we are living with such universal correlations, on large scales where the nonlinearity is 
fairly weak, keeping the correlation between different modes to ensure non-zero bispectrum we can 
make an assumption to neglect the correlation between the amplitude and the phase of a single mode, an 
interesting conclusion emerges. With the assumption, the ensemble average 
of bispectrum can be decomposed into
\begin{equation}
B({\bf k}_1, {\bf k}_2, {\bf k}_3) \propto \langle |\delta_1| |\delta_2| |\delta_3| \rangle \int e^{{\it i}\Theta} 
p(\Theta)d\Theta\ ,
\end{equation}
in which $p({\Theta})$ is the PDF of the phase sum $\Theta=\theta_1+\theta_2-\theta_3$. Since 
the bispectrum is real and $\delta({\bf k})$ is Fourier transform of a real function, 
$p(\Theta)$ shall be symmetric to $\Theta=0$ and can be expanded as 
$p(\Theta)=c_0+\sum_n c_n\cos(n\Theta)$. Keeping only the first two terms, we
recover the essence of the results of Matsubara (2003)
\begin{equation}
p(\Theta)\propto \left( 1+const.\cdot \cos\Theta\right) \ .
\end{equation}

In redshift space, modifications occur in both amplitudes and phases, 
\begin{equation}
\delta_s({\bf k})=|\delta_s|e^{{\it i}\theta_s} = \alpha({\bf k}) |\delta({\bf k})| e^{{\it i} (\theta+\Delta\theta)}\ ,
\end{equation}
in which the amplification factor $\alpha({\bf k},\mu)$ is real and phase shift is
$\Delta \theta({\bf k}) \equiv \theta_s({\bf k})-\theta({\bf k})$. Thus the bispectrum in redshift space
can be written as
\begin{equation}
B_s({\bf k}_1, {\bf k}_2, {\bf k}_3=-{\bf k}_1-{\bf k}_2)=\langle \alpha_1 \alpha_2 \alpha_3 |\delta_1| 
|\delta_2| |\delta_3| e^{{\it i} (\theta_1+\theta_2-\theta_3)} e^{{\it i} (\Delta\theta_1+\Delta\theta_2-\Delta\theta_3)}
\rangle\ .
\end{equation}
According to our observation of the Virgo and the VLS simulations, $\alpha({\bf k})=\alpha(k, \mu)$ is 
deterministic and can be well approximated by $\alpha=(1+\beta \mu^2)/[1+(k\mu\sigma_v)^2/2]$ as in 
Eq.~2, hence the product of amplitude amplification factors in Eq.~14 can be placed outside
the ensemble average. 

If we assume at a given ${\bf k}$, $\Delta\theta$ is fixed regardless of the original 
stochastics, $\Delta\Theta=\Delta\theta_1+\Delta\theta_2-\Delta\theta_3$ will be a constant so that 
$B_s=\alpha_1\alpha_2\alpha_3 B e^{{\it i} \Delta\Theta}$.
$\Delta\Theta$ has to be $n\pi \ (n=0,\pm 1,\pm 2 ...)$ as $B_s$ is real, so
\begin{equation}
B_s=\alpha_1\alpha_2\alpha_3 B\ .
\end{equation}
It is hard to conjecture there is cosmic conspiracy to satisfy $\Delta\Theta=n\pi$ for arbitrary
$({\bf k}_1, {\bf k}_2,{\bf k}_3)$ triplet, the only possibility is for any mode 
$\Delta\theta=n\pi$, which is obviously impossible as implied by Eq.~7 \& 8. The immediate 
conclusion is the phase shift resulted from redshift distortion has dependence on the phase in
real space and therefore is stochastic.

In N-body simulations we found that although phases no matter in real space or in 
redshift space are uniformly distributed, the distribution of phase shift $p(\Delta\theta)$ is 
not uniform as shown in Fig.~(1). In general $p(\Delta\theta)$ depends on $(k,\mu)$ and is peaked 
at center $\Delta\theta=0$. With the decreasing of $\mu$ and $k$, the dispersion of the distribution is 
becoming smaller and more modes concentrate on the vicinity of $\Delta\theta=0$. This is easy to 
understand since those modes around the plane perpendicular to LOS 
suffers little from redshift distortion (Eq.~7) while on large scales the deviation to 
linear redshift distortion theory of Eq.~9 becomes small. We examined that in simulations phases of
modes on the plane $\mu=0$ are unshifted.

\begin{figure}
\resizebox{\hsize}{!}{
\includegraphics{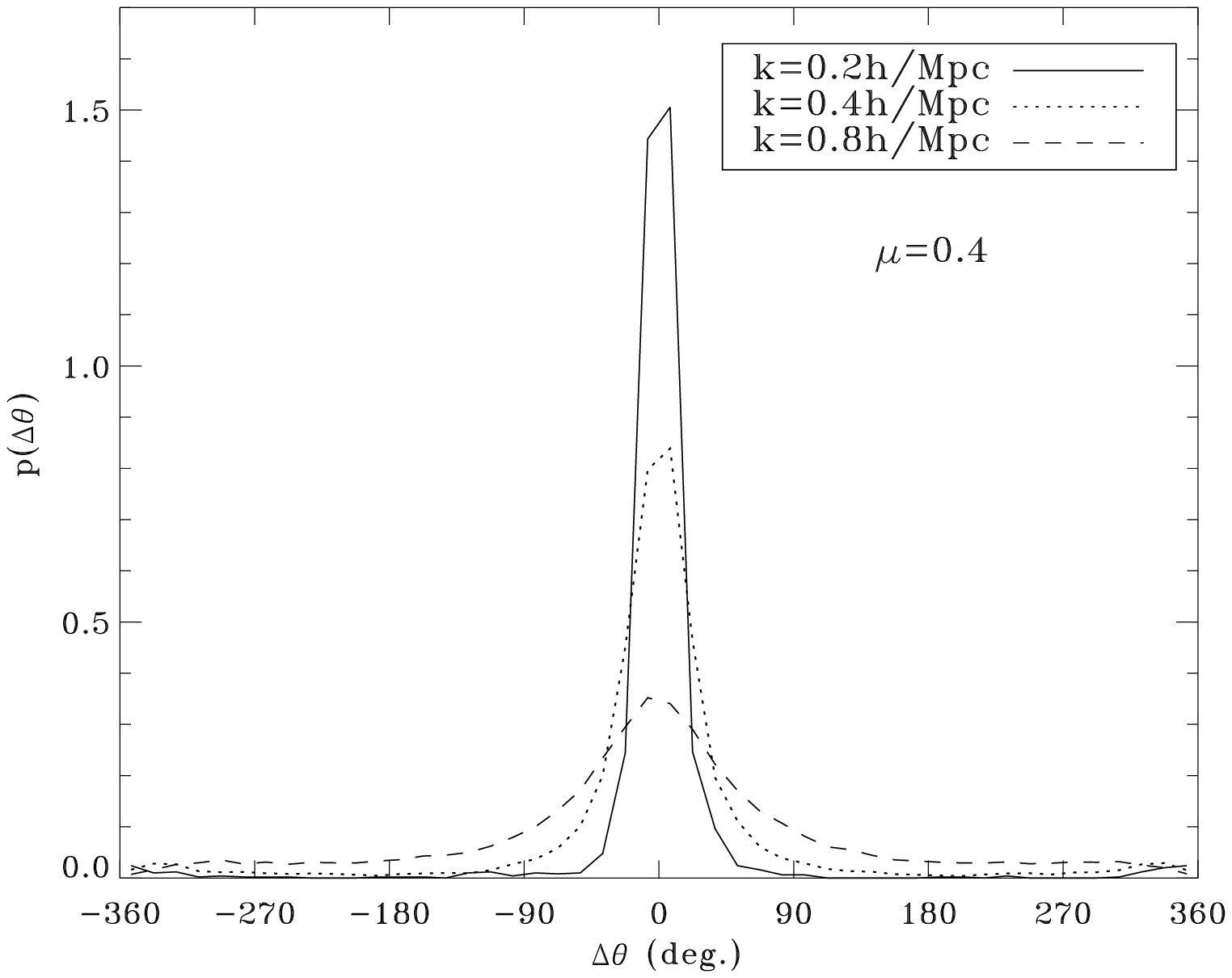}
\includegraphics{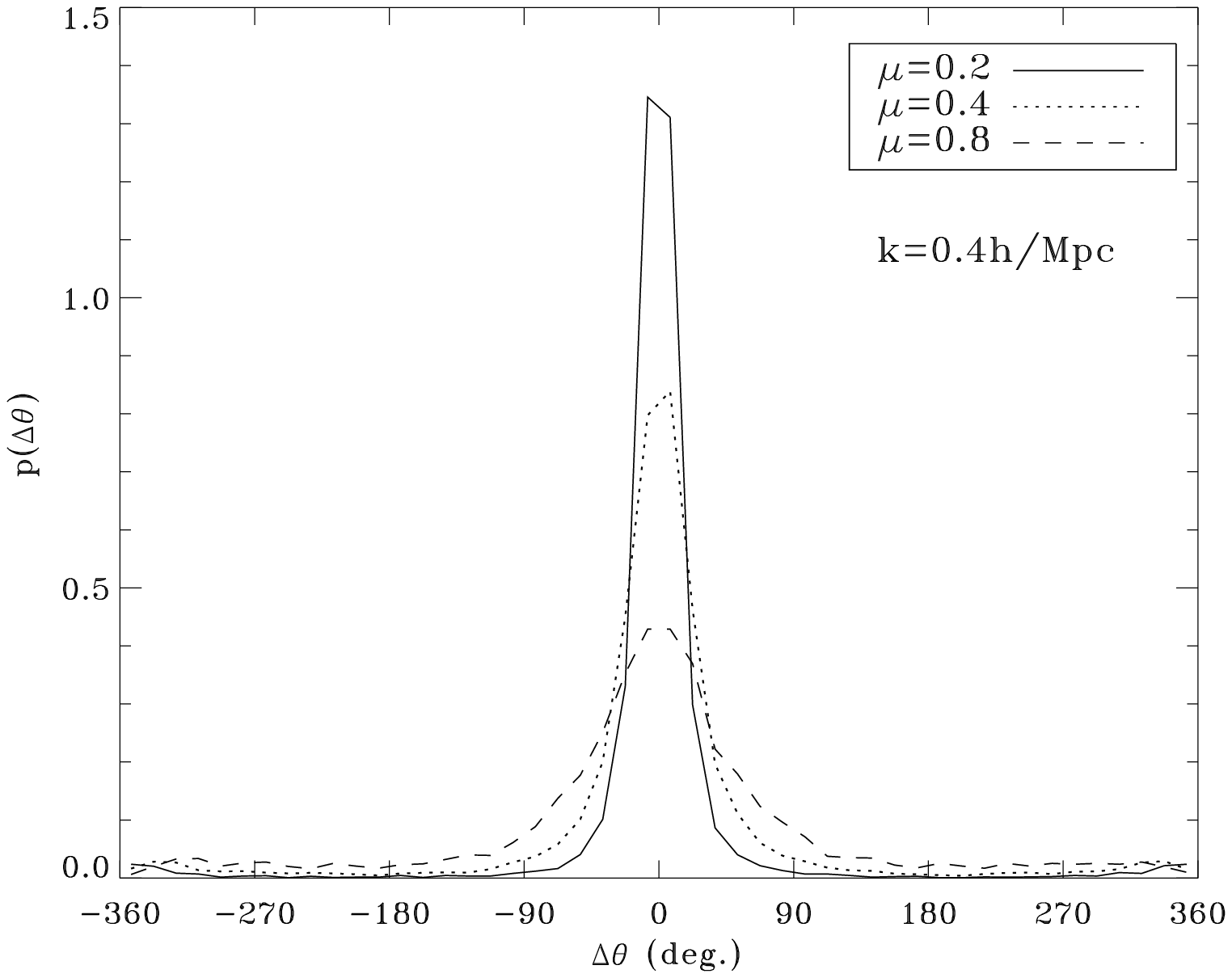}
}
\caption{Probability distribution functions of phase shift $\Delta\theta$ resulted from redshift 
distortion in N-body simulations. The left panel plots the $p(\Delta \theta)$ on different $k$ scales 
at fixed $\mu={\bf k}\cdot {\hat z}/k$, and in the right panel we show $p(\Delta \theta)$ at 
variant $\mu$ at $k=0.4h \rm{Mpc^{-1}}$.}
\end{figure}

\subsection{Toy models for bispectrum in redshift space}
In attempt to understand the redshift distortion of bispectrum, as a crude approximation in light of
the spirit of deriving Eq.~12 we assume that $\Delta\Theta$ is uncorrelated with $|\delta_{1,2,3}|$ 
and $\theta_{1,2,3}$, which is possible since $\Delta\Theta$ is a sum of three random variables.
Eq.~14 then becomes 
$B_s=B\times \alpha_1 \alpha_2 \alpha_3 \int e^{{\it i} \Delta\Theta}p(\Delta\Theta)d\Delta\Theta$, and
$p(\Delta\Theta)\propto (1+const. \cdot \cos\Delta\Theta)$ if high order terms are ignored. With a further
assumption that $\Delta\theta_{1,2,3}$ is independent of each other, we have
\begin{equation}
\begin{aligned}
B_s({\bf k}_1, {\bf k}_2, {\bf k}_3)=&B({\bf k}_1, {\bf k}_2, {\bf k}_3)
\times \alpha_1 \alpha_2 \alpha_3 \\ \times
&\int e^{{\it i} \Delta\theta_1}p(\Delta\theta_1)d\Delta\theta_1
\int e^{{\it i} \Delta\theta_2}p(\Delta\theta_2)d\Delta\theta_2
\int e^{{\it i} \Delta\theta_3}p(\Delta\theta_3)d\Delta\theta_3\ ,
\end{aligned}
\end{equation}
from which trivially we again have features of $p(\Delta\theta)$ that it is symmetric to
$\Delta\theta=0$ seen in Fig.~1 and $\propto (1+const.\cdot \cos\Delta\theta)$.
Note that here $\alpha$ and $\Delta\theta$ are functions of two variables $(k,\mu)$.
Eq.~16 is quite similar to the phenomenological model of Scoccimarro et al. (1999), and 
it serves as a theoretical support to their {\em ansatz}. 

In practice often it is simpler to measure of the bispectrum averaged over all possible combinations
of wave vectors $({\bf k}_1, {\bf k}_2, {\bf k}_3=-{\bf k}_1-{\bf k}_2)$ with fixed $k_1=|{\bf k}_1|$,
$k_2=|{\bf k}_2|$ and the angle $\psi={\bf k}_1\cdot{\bf k}_2/(k_1k_2)$. In real space it is 
the isotropic bispectrum $B(k_1,k_2, k_3)=B({\bf k}_1, {\bf k}_2, {\bf k}_3)$, in redshift space
due to the anisotropy of redshift distortion, the angular averaged with respect to LOS is the monopole of
true bispectrum $B_s(k_1, k_2, \psi, \mu_1, \mu_2, \mu_3)$. In that we denote
the azimuthal angle of ${\bf k}_2$ about ${\bf k}_1$ with $\varphi$, and 
\begin{equation}
\begin{aligned}
\mu_1&=\mu={\bf k}_1 \cdot {\hat{z}}/k_1\ ,\\
\mu_2&=\mu\cos\psi-\sqrt{1-\mu^2}\sin\psi\cos\varphi\ , \\
\mu_3&=-\frac{k_1}{k_3}\mu-\frac{k_2}{k_3}\mu_2\ ,
\end{aligned}
\end{equation}
the monopole of bispectrum in redshift space is
\begin{equation}
B^{(0)}_s(k_1, k_2, \psi)=\frac{1}{4\pi} \int_{-1}^{1} d\mu \int_0^{2\pi} d\varphi B_s(k_1, k_2, \psi,\mu, \varphi)\ .
\end{equation}
On large scales where the nonlinearity is weak and phases shifts are close to zero (Eq.~9 and Fig.~1), 
we can approximate the modulation by peculiar velocity with Eq.~15, 
\begin{equation}
\frac{B_s^{(0)}}{B}=\frac{1}{4\pi}\int_{-1}^{1}\int_0^{2\pi}\alpha(k_1, \mu)
\alpha(k_2,\mu,\psi,\varphi)\alpha(k_3,\mu,\psi,\varphi)d\varphi\ d\mu.
\end{equation}

We measured angular averaged power spectrum and bispectrum both in real space and redshift space for 
the Virgo simulation and eight subsets of the VLS simulation which are generated by dividing the VLS 
simulation into independent cubes of half of the original box size. $\sigma_v=4.87$ is fitted from 
power spectra with Eq.~2, and then is inserted into Eq.~19 as our toy model for $B_s^{(0)}$.
\begin{figure}
\begin{center}
\includegraphics[height=70mm, width=55mm, angle=-90]{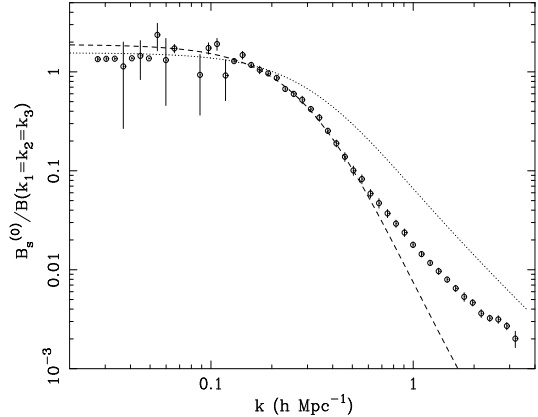}
\end{center}
\caption{$B^{(0)}_s/B$ of equilateral triangles. Dotted line is the prediction of Eq.~19, 
and dash line is the empirical model of Eq.~20. Points are averaged over measurement of the 
Virgo simulation and the eight subsets of the VLS simulation. Error bars are the $1\sigma$ scatter.}
\end{figure}
In Fig.~2 it is interesting to observe the remarkable agreement of our toy model based on phase information
with simulations for equilateral triangles for $k < 0.2h \rm{Mpc^{-1}}$ beyond which tree-level 
perturbation theory and the second order Lagrangian theory also breaks (Scoccimarro 2000).
As $k$ grows, phase shift is no longer negligible and our toy model fails.

\begin{figure}
\resizebox{\hsize}{!}{\includegraphics[angle=-90]{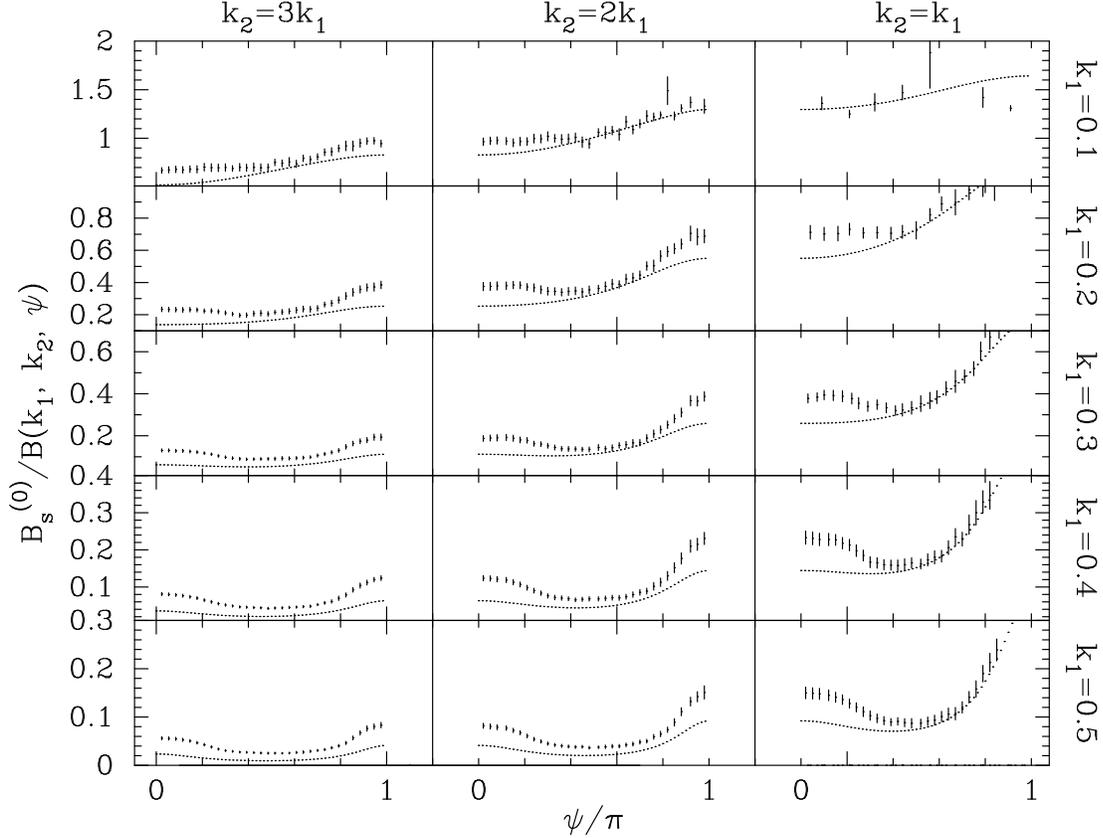}}
\caption{$B^{(0)}_s/B(k_1, k_2, \psi)$ of different configurations. The dotted line is our modified
toy model Eq.~20, see Fig.~2 for other keys.}
\end{figure}

As a next level advanced than the simple prescription of Eq.~19, a better approximation may be 
achieved through integration of Eq.~16 instead of the simple formula of Eq.~19 if we know 
$p(\Delta\theta)$ which carries important information about the redshift distortion and numerically 
contains more dependence on $(k,\mu)$ (Fig.~1).
Because we are not equipped with a whole battery of many simulations in very large box, so far in this
work we are not developing functional expression to fit $p(\Delta\Theta)$'s dependence on $(k,\mu)$ 
and $\sigma_v$. Nevertheless, we can parameterize the toy model in a simple way to improve its 
performance just as in Scoccimarro et al. (1999), e.g.
\begin{equation}
\frac{B_s^{(0)}}{B}=A_1\left[ \frac{1}{4\pi} \int_{-1}^{1}\int_0^{2\pi}\alpha(k_1, \mu)
\alpha(k_2,\mu,\psi,\varphi)\alpha(k_3,\mu,\psi,\varphi)d\varphi\ d\mu\right]^{A_2}\ , 
\end{equation}
with two fitting parameters $A_{1,2}$. An eye-ball examination with $A_1=1/1.15$ and $A_2=1.75$ does
improve agreement up to scale $k \sim 0.5h\rm{Mpc^{-1}}$ for equilateral triangles (Fig.~2). We 
then compare the modified toy model for other configurations with simulations in Fig.~3. We see that 
for isosceles with $\psi > \sim 90^\circ$ the agreement of our modified toy model with simulations is 
very impressive too, although for those very tilted configuration with $k_2/k_1\ge 2$ differences are 
huge. We are not intending to give accurate fit, we are just demonstrating the possibility of 
constructing a good approximation model with the simple prescription of Eq.~19, and, in the mean time 
showing the discrepancy due to our rough handling with phase shift.


\section{Goodness of Plane-Parallel Approximation to Radial Distortion}

So far we are working with plane-parallel approximation. As we have addressed in section 1, such 
approximation is appropriate when the angle subtended by the sample is small. When dealing with 
large samples, especially all sky surveys, radial distortion should be taken into account. To 
investigate the goodness of the plane-parallel approximation to the exact 
radial distortion, we are to compare a set of power spectra and bispectra in redshift space 
in the parallel approximation and the radial distortion.

\subsection{Monopoles of Power Spectrum and Bispectrum}
It has been shown by Scoccimarro (2000) that the monopoles of power spectrum and bispectrum of 
an all sky survey, in the radial and the plane-parallel distortions are consistent with each other
within error bars. For the power spectrum the difference is within $\sim 10\%$ and for the 
bispectrum is within a few percents. 

To generate samples in redshift space in radial distortion, we strictly follow Eq.~5 with
$\hat{z}$ defined by the unit vector pointing from the observer to the point of sample, i.e. 
$\hat{z}$ is simply the unit position vector $\hat{\bf r}$. Cartesian coordinates $r_{i=1,2,3}$ 
of any point in our samples are within $[0,L]h^{-1}$Mpc, $L=239.5$. In this 
subsection, we placed the observer at $(0.5\sqrt{2}L, 0.5\sqrt{2}L,  -100)h^{-1}$Mpc, 
the largest angle of the sample opening to the observer is $118.88^\circ$. Then points in samples
are transformed into the coordinate system in which the observer is sitting at the origin point. 
For every data point we have redshift from Hubble flow $z_{Hub}$ and peculiar velocity induced $z_{pec}$, 
the comoving distance in redshift space $s$ is then calculated from the final redshift 
$z=(1+z_{Hub})(1+z_{pec})-1$ while its orientation does not change.
  
In Fig.~4, we plotted the ratio of the monopoles of power spectra in radial distortion and the
parallel approximation, $P^{(0)}_r/P^{(0)}_p$ where the subscript ``r'' means 
radial distortion, ``p'' refers to parallel approximation and all polyspectra are in redshift
space if not specified hereafter. Monopole of power spectrum is
defined as the angular average of the anisotropic $P({\bf k})$.
Comparison of the monopoles of reduced bispectrum is illustrated in Fig.~5. Note that the monopole
of reduce bispectrum is given by
\begin{equation}
Q^{(0)}(k_1,k_2,\psi) = \frac{B^{(0)}(k_1,k_2,\psi)}{P^{(0)}(k_1) P^{(0)}(k_2) + P^{(0)}(k_2) P^{(0)}(k_3)+
P^{(0)}(k_3) P^{(0)}(k_1) }
\end{equation}

\begin{figure}
\begin{center}
\includegraphics[width=55mm,height=70mm,angle=-90]{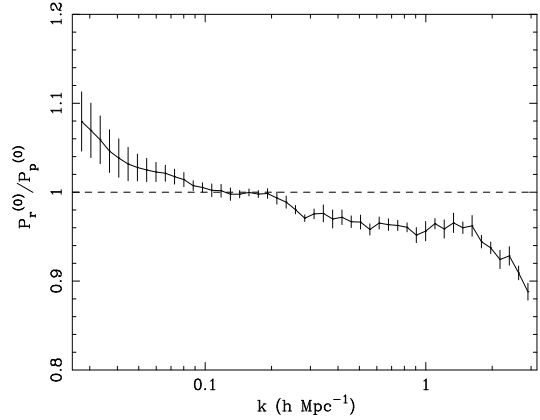}
\end{center}
\caption{Ratio of the monopoles of power spectra in radial distortion and parallel approximation. 
The sample is at the distance of $100h^{-1}$Mpc.}
\end{figure}

\begin{figure}
\begin{center}
\resizebox{\hsize}{!}{\includegraphics[angle=-90]{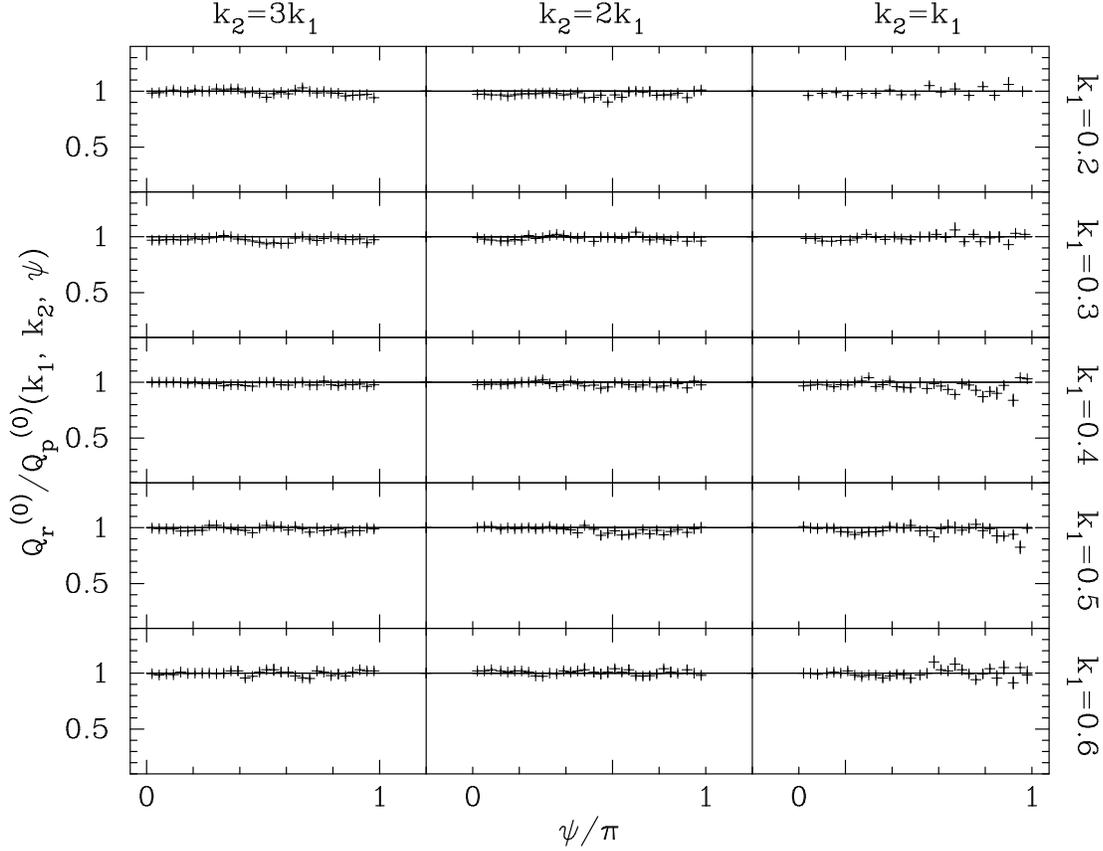}}
\end{center}
\caption{$Q^{(0)}_r/Q^{(0)}_p(k_1, k_2, \psi)$ for triangles of $k_1=0.2,0.3,0.4,0.5,0.6h \rm{Mpc^{-1}}$ 
and $k_2/k_1=1,2,3$. The observer is at the distance of $100 h^{-1}$Mpc.}
\end{figure}

We can see from Fig.~4 that the differences between monopoles of power spectrum in the radial and the
plane-parallel distortion are within $10\%$, while the differences for bispectra are of a few 
percents (Fig.~5). These results confirm the assertion of Zaroubi \& Hoffman (1996) and 
Scoccimarro (2000). As shown in Fig.~4, the power in parallel approximation on large scales 
$k< 0.1 h \rm{Mpc^{-1}}$ is smaller than that under radial distortion. On the other hand, on small 
scales $k > 0.2 h\rm{Mpc^{-1}}$, the plane-parallel distorted power spectrum is larger. This is because 
in the plane-parallel approximation there is no distortion effect along directions perpendicular to 
the fixed LOS, the actual redshift distortion is radial and the LOS is not fixed. Distortion is more 
severe to structure clumps so that there are more effects of boost at large scales and suppression 
at small scales on the clustering strength.

\subsection{Power Spectrum and Bispectrum on the Plane of $k_z=0$}
We have seen that there is systematic bias caused by the parallel approximation to the radial distortion 
in monopoles of power spectrum and bispectrum though very weak. In this subsection we set about to 
explore the behavior of the polyspectra on the plane of $k_z=0$ which are regarded as estimation of 
the real space quantities in the plane-parallel approximation. It is easy to understand that on large 
scales where the opening angle to the observer is wide so the lines of sight can not be approximated by 
parallel lines, such deviation is not separable in Fourier transform which is a global operation. Only 
when the opening angle of the sample window is small we are safe. In order to check the influence of the 
opening angle of sample to the method of using polyspectra of $k_z=0$ as estimation in real space, we 
created different samples in redshift space from simulations by placing the observer at distances of 
100, 200 and 1000$h^{-1}$Mpc respectively.

\begin{figure}
\begin{center}
\includegraphics[width=55mm,height=70mm,angle=-90]{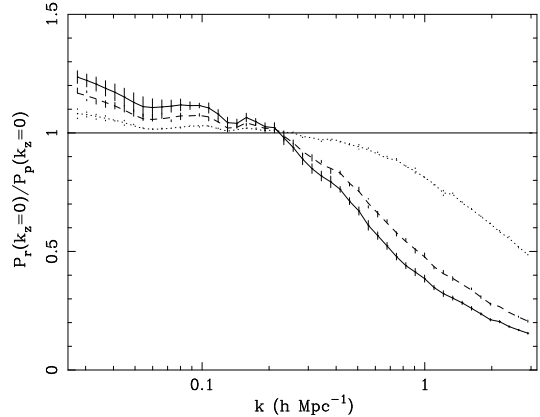}
\end{center}
\caption{Ratios of power spectra for radial and parallel distortion on the plane of $k_z=0$. The solid 
line, dash line and dot line are for samples at distances of $100 h\rm{Mpc^{-1}}$, $200 h\rm{Mpc^{-1}}$ 
and $1000 h\rm{Mpc^{-1}}$ to the observer respectively.}
\end{figure}

\begin{figure}
\begin{center}
\resizebox{\hsize}{!}{\includegraphics[angle=-90]{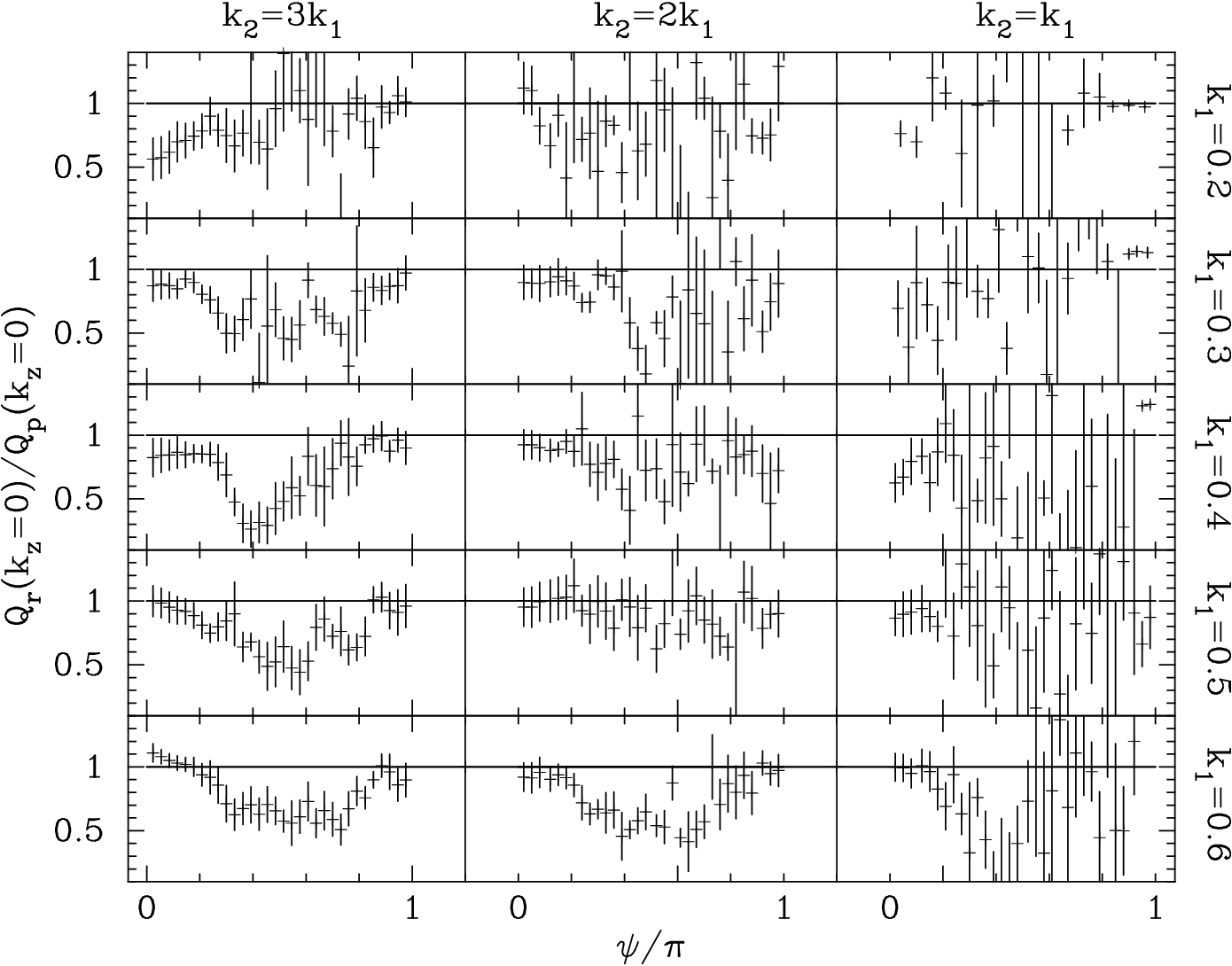}}
\end{center}
\caption{Ratios of reduced bispectra for radial and parallel distortion on the plane of $k_z=0$. Observer 
is located at the distance of $100h^{-1}$Mpc.}
\end{figure}

\begin{figure}
\begin{center}
\resizebox{\hsize}{!}{\includegraphics[angle=-90]{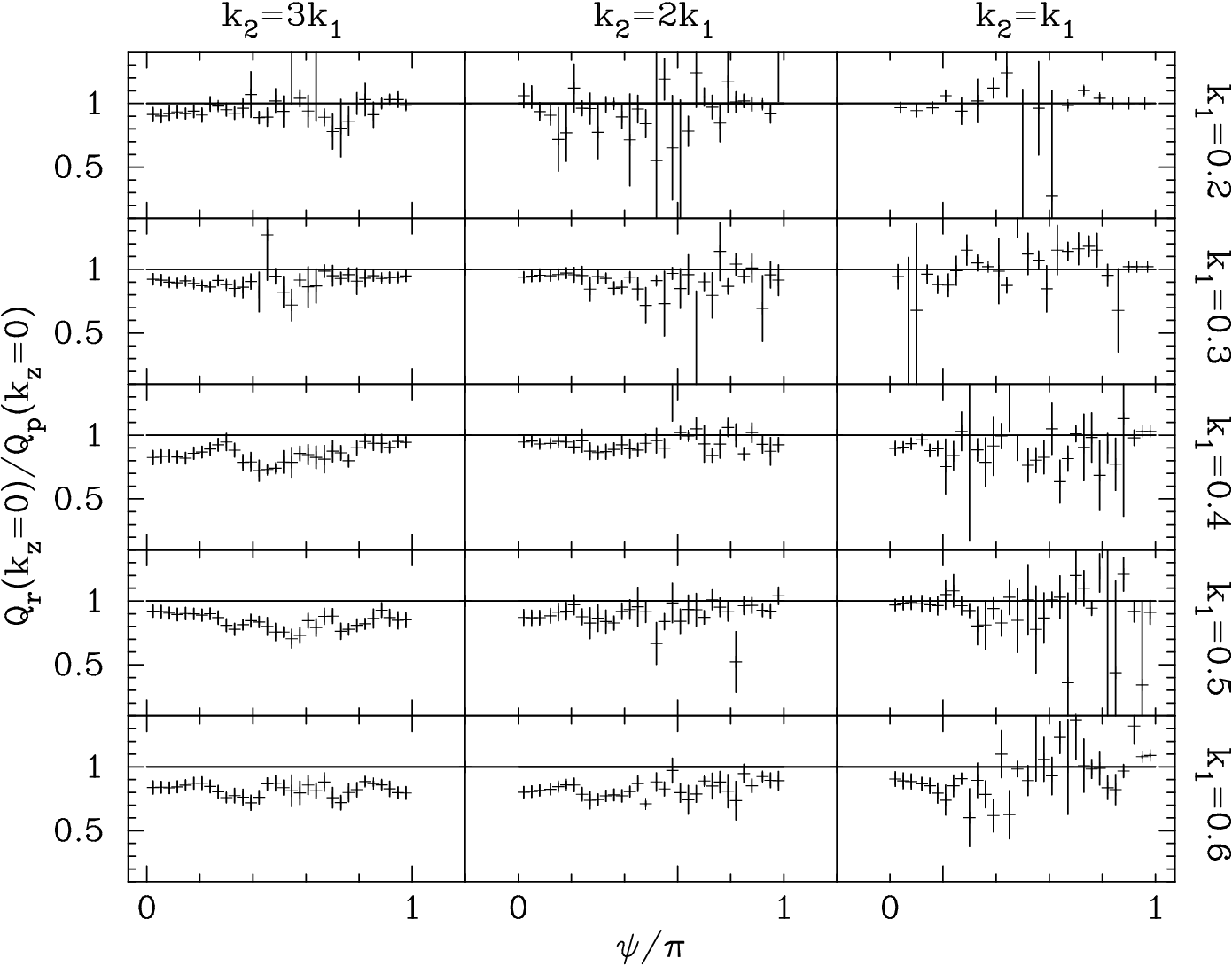}}
\end{center}
\caption{Ratios of reduced bispectra for radial and parallel distortion on the plane of $k_z=0$. Observer 
is located at the distance of $1000h^{-1}$Mpc.}
\end{figure}

Fig.~6 shows that when $k$ is larger than $0.2h{\rm Mpc}^{-1}$, under the plane-parallel approximation,  
$P_p$ can grows from $\sim 10\%$ to more than $50\%$ larger than $P_r$; when $k$ is smaller than 
$0.2 h\rm{Mpc^{-1}}$, $P_p$ could still be $>10\%$ smaller than $P_r$. Meanwhile, as we expected, 
with increment of the distance of the sample to observer, i.e. the opening angle getting smaller, such 
difference systematically decrease. 

However, even when the distance is 1$h^{-1}$Gpc which corresponds 
to an opening angle as small as $\sim 20^\circ$, at $k=3h \rm{Mpc^{-1}}$ the systematical bias persists
around $50\%$ if one is interested in the highly nonlinear regime, meanwhile the bias
is also $>10\%$ for $k<0.03h \rm{Mpc^{-1}}$ which prevents us from having measurements in {\em precision}.
It is a bad news for most redshift surveys, such as the 2dFGRS redshift survey and the 
SDSS of which subtended angles are much larger than $20^\circ$. Furthermore, as seen in Fig.~7 and 
Fig.~8, there are similar systematic biases in reduced bispectrum measured at $k_z=0$ to the real space 
bispectrum if we adopt the plane-parallel approximation, the biases not only attenuate the amplitudes
but also the configuration dependence.

Therefore, it is hopeless to accurately estimate the real space polyspectra on large scales with
the convenient method of using what we measured on the plane $k_z=0$ in redshift space if no other
special skills are applied to compensate the systematical bias. One of the possible application of 
the method is to estimate small scale polyspectra in real space by splitting the wide angle sample 
into many patches of very small angles in compromise with the scales of interests, and later collecting
measurements in these small patches for analysis in precision, which, of course, needs careful numerical 
experiments to decide the angular widths of those patches (e.g. Park et al. 1994).

\section{Summary and Discussion}
In the exploration of the characteristics of redshift distortion in Fourier space, we begin with 
research on phases of Fourier modes which are rarely noticed in previous works. On very large scales 
where linear theory applies, phases are not changed by redshift distortion, we conclude that 
phase shift $\Delta\theta$ is a nonlinear effect. Then we find that the phase shift by redshift 
distortion is not deterministic, it has a distribution function $p(\Delta\theta)$ symmetric to 
its peak at $\Delta\theta=0$, and qualitatively proportional to $(1+const.\cdot\cos\Delta\theta)$. 
The distribution function has strong dependence on the wave length $k$ and the orientation $\mu$ of 
the wave vector to the LOS. More modes are populated around $\Delta\theta=0$ for small $k$ and 
small $\mu$.

Limited by the sizes of the available Virgo and VLS simulations and computing resources, we can not 
obtain reliable $p(\Delta\theta)$ on scales of $k<0.1h \rm{Mpc^{-1}}$ to build empirical model for 
the distribution function. Although in principle it is possible to calculate the phase shift in
perturbation theory, it is unclear how to generate the PDF of the phase shift in perturbation
approach, which we leave to future work.
Instead we make assumptions on the phase shift at different levels in effort 
to depict the redshift distortion on bispectrum. We discovered that our very crude consideration does 
provide us with a toy model (Eq.~19) for the monopole of bispectrum in redshift space which is 
in good agreement with numerical simulations on scales $k<0.2h \rm{Mpc^{-1}}$, and
probably can serve as a basis for extended phenomenological models such as Eq.~20 and the model in
Scoccimarro et al. (1999). If in future an approximation for $p(\Delta\theta)$ analogous to the modeling 
of amplitude change like Eq.~(2) can be set with help of large simulations, it is promising that 
integration of Eq.~16 will offer us with greatly improved model for the monopole of bispectrum in 
redshift space.

On non-linear scales, the plane-wave components of density fluctuation do not evolve independently, 
interaction among difference modes leads to generation of coupled phases. Therefore, significant phase 
correlation between different modes is expected. As the Fourier phases of a random Gaussian field are 
randomly distributed, phase correlation, if any, should characterizes the non-Gaussianity of density 
fields (Matsubara 2003). Such correlation would have important effects on the bispectrum, which is the 
lowest order statistics of non-Gaussianity. The importance of connection between phase correlation and 
non-Gaussianity is doubtless. By using a perturbative quadratic model, Watts \& Coles (2003) have shown 
for the first time how phase association give rise to non-vanishing bispectrum and three point function 
in nonlinear processes. The arrangement of phases of modes determines weather such non-Gaussian 
descriptors are zero or not, but the magnitude of non-Gaussianity is determined by the magnitude of 
Fourier modes (Watts \& Coles 2003). A relationship between phase correlations and the hierarchy of 
polyspectra in the Fourier space has been addressed by Matsubara (2003). Also, a numerical test on the 
bispectrum has been given. However, many details of this issue such as how phase correlation functions 
with redshift distortion and bias, remain unknown.

The connection between polyspectra and phase information may render us much more useful information, 
while our knowledge of it is till limited. It is worthwhile for us to tackle the problem in more 
details by taking into account of both the phase shift of individual Fourier mode which is presented
in this paper, and the correlation between phase shifts of different modes which importance is of 
priority in understanding the redshift distortion on bispectrum in strong nonlinear regime. As a first
step, the simplest consideration on phase shift already has guided us in producing a good model, future
inclusion of the PDF and the correlation between phase shift will shed light on the redshift distortion
in Fourier space. 
 
In the second part of our report, we explored the goodness of the plane-parallel approximation to the 
actual radial redshift distortion in Fourier analysis. The first examination is exerted to the monopoles 
of power spectrum and bispectrum. Our measurements with the Virgo and the VLS simulation are consistent 
with the results of Scoccimarro et al. (2000) that such deviation due to adoption of parallel 
approximation is $\sim 10\%$ for the monopole of power spectrum and a few percents for bispectrum. 
Note that the curve in Fig.~4 shows a systematic trend toward high $k$. 
Then, we moved to check the reliability of the approach which uses polyspectra measured on $k_z=0$ 
plane as estimation in real space, which is true in plane-parallel approximation. We find that 
plane-parallel approximation to radial distortion brings in serious statistics systematic biases in 
this method. On large scales the ``real space'' power spectrum from measurements on $k_z=0$ plane can 
be overestimated for more than $10\%$, and it is hugely underestimated more than $50\%$ on small scales 
$k> \sim 3h \rm{Mpc^{-1}}$ (Fig.~6). We also find that besides the amplitude of reduced bispectrum 
changes, the configuration dependence suffers from the badness of parallel approximation too 
(Fig.~8 \& 9). According to our simulation, such deviation is persistent even the opening angle of 
the sample to the observer is as small as $20^\circ$. These discoveries pose questions on those 
claims based on the ``real space'' polyspectra obtained in this way.

So as long as we are working with monopoles of power spectrum and bispectrum, it is secure to model 
redshift distortion with plane-parallel approximation. While if one is going to estimate real space 
power spectrum and bispectrum with measurements on plane $k_z=0$, one has to be careful to consider 
the caveats of the parallel approximation. Actually a more natural treatment is to decomposed the 
density contrast in redshift space with spherical harmonics and spherical Bessel functions 
(Heavens \& Taylor 1995; Ballinger, Heavens \& Taylor 1995; Fisher et al. 1995; 
Szalay, Matsubara \& Landy 1998; Percival et al. 2004; Percival 2004; Szapudi 2004).

\begin{acknowledgements}
Yan-Chuan Cai thanks Prof. Long-Long Feng for support of the work and helpful comments. This work was 
supported by the National Science Foundation of China through grant NSFC 10373012. Jun Pan appreciates 
the support by the One-Hundred-Talents program. The simulations in this 
paper were carried out by the Virgo Supercomputing Consortium using computers based at 
Computing Center of the Max-Planck Society in Garching and at the Edinburgh Parallel Computing 
Center. The data are publicly available at www.mpa-garching.mpg.de/NumCos.
\end{acknowledgements}

\label{lastpage}
\end{document}